\documentclass[a4paper,10pt]{article}
\usepackage[utf8]{inputenc}

\title{  Deformation of the Wheeler-DeWitt Equation}
\author{Mir Faizal \\Department of Physics and Astronomy, \\  University of Waterloo,   Waterloo,\\
Ontario N2L 3G1, Canada 
}
\date{}
\begin{document}

\maketitle

\begin{abstract}
In this paper, we will analyse the consequences of deforming the canonical commutation 
relations consistent with the existence of a minimum length 
and a maximum momentum. We first generalize the deformation of first quantized canonical commutation relation 
to second quantized canonical commutation relation. Thus, we arrive at a modified version of second quantization.  A
modified  Wheeler-DeWitt equation will be constructed by using this deformed  second quantized canonical commutation relation.
Finally, we demonstrate that in this modified theory the big bang singularity gets naturally avoided. 
\end{abstract}
\section{Introduction}
In any approach to quantum gravity, the classical picture of spacetime is expect to break down. This is 
because at Planck scale the fluctuations in the geometry are expected to become of order unity. Thus, 
the picture of spacetime as a differential manifold cannot hold below Planck scale. In fact, 
many studies on physics of the black holes have suggested that all quantum gravity theories should have a
minimum measurable length of the order of the Planck length \cite{z4}-\cite{z5}.
The string theory also come naturally equipped  with a minimum length\cite{z2}-\cite{2z}. In fact,  in loop quantum gravity  
the existence of minimum length 
 turns big bang into a big bounce \cite{z1}.
 It may be noted that the existence of a minimum lent is not consistent with the 
 conventional Heisenberg uncertainty principle. This is because according to the conventional 
 Heisenberg uncertainty principle, the minimum measurable length is actually zero. 
 To remove this inconsistency, the conventional Heisenberg principle is modified \cite{1}-\cite{54}. 
 The resultant uncertainty principle is called the generalized uncertainty principle. 
The modification of the uncertainty principle naturally leads to a  modification of the Heisenberg algebra. 
In this new Heisenberg algebra commutation relations between position
and momentum operators contain momentum dependent factors. It may be noted that the implications of this modified 
uncertainty principle for quantum field theory have also been studied \cite{n9}-\cite{n}.

In doubly special special relativity theories, the Planck energy like the speed of light is a
invariant quantity in doubly special special relativity. This is incorporated by 
 a modification  of the Heisenberg  algebra \cite{2}-\cite{3}. 
 A consequence of this modification is that, doubly special relativity come naturally equipped with a maximum measurable momentum. 
A modified version  of  general relativity called the Gravity's Rainbow has been constructed \cite{n1}-\cite{n2}. 
Gravity's Rainbow
both velocity of light and the Planck energy are again 
invariant quantities. 
Thus, two different deformations of Heisenberg algebra have been studied. In fact, a
algebra has also been constructed which is consistent with the existence of both a minimum length and  a maximum momentum \cite{n4}-\cite{n5}.
The transition rate of ultra cold neutrons in gravitational field
has been analysed using this deformed algebra \cite{n6}. 
In fact,  the modification to the Lamb shift and Landau levels have
also been analysed in this deformed algebra \cite{n7}. 

The relation between the quantum mechanical commutators and the  classical Poisson brackets has also been used to study modification to the 
Friedman-Robertson-Walker cosmology \cite{fwrw}. In this paper, we will deform the second quantized commutating relations and study the 
effect of the deformation on Wheeler-DeWitt equation. The solutions to the Wheeler-DeWitt equation gives us the wave function of the universes
 \cite{DeWitt67}-\cite{Wheeler57}. As the wave  function of the universe describes  the 
quantum state of the universe, so all the 
physical information about the universe can be extracted from it \cite{Hartle83}-\cite{Hawking}. This wave function can also be obtained 
by taking a sum over all geometries and field configurations which match with a particular field configuration 
at a spatial section of the spacetime. 
This approach is called the the Hartle-Hawking
no-boundary proposal. 
In this approach a Wick rotation to Euclidean time makes this integral well defined. 
This wave function of the universe can also be viewed as a solution of the Wheeler-DeWitt equation \cite{DeWitt67}-\cite{Wheeler57}.
The Wheeler-DeWitt equation is a second quantized equation which can be interpreted as the Schroedinger's equation for gravity. 
However, there is no time in the Wheeler-DeWitt equation because it has to satisfy the time invariance which is required by 
 general relativity \cite{Hawking81}. In this paper we will analyse the implications of modifying the Wheeler-DeWitt equation, by 
 using a deformed version of second quantized canonical commutator. Usually, the generalized uncertainty is studied as a implication 
 of some quantum gravitational effect. However, in this paper, we will reverse this and study the implications of generalized uncertainty
 on quantum gravity. 
\section{ Deformed Heisenberg Algebra}
In this section, we will first review the deformation of the 
Heisenberg consistent with the existence of minimum  length and maximum momentum. 
Then we will construct a second quantized version of this algebra. We let $\ell_{Pl}\approx 10^{-35}~m$ be the Planck length, 
$M_{Pl}$ be the Planck mass, $M_{Pl} c^2 \approx 10^{19}~GeV$ be the Planck energy, and 
$\beta_0$ be a constant of order unity. 
Now the existence of a minimum length is consistent with
the following algebra   $[x^i, p_j ] = 
i [\delta_{j}^i + \beta p^2 \delta_{j}^i + 2 \beta p^i p_j]$, and the existence of a maximum momentum is consistent with the following algebra  
$ [x^i, p_j] =
i [1 - \tilde \beta |p| \delta_{j}^i + \tilde \beta p^i p_j]$, where $\beta = \beta_0 \ell_{Pl}/ \hbar$ and 
$\tilde \beta = \ell_{Pl}$. 
Both these deformations of Heisenberg algebra have been combined into a single  algebra \cite{n4}-\cite{n5}
\begin{equation}
 [x^i, p_j] = i \left[  \delta_{j}^i - \alpha ||p|| \delta_{j}^i + \alpha ||p||^{-1} p^i p_j 
  + \alpha^2 p^2 \delta_{j}^i + 3 \alpha^2 p^i p_j\right],
\end{equation}
with $\alpha = {\alpha_0}/{M_{Pl}c} = {\alpha_0 \ell_{Pl}}/{\hbar}$. It may be noted that the norm of the momentum 
is defined as 
\begin{equation}
 || p || = \sqrt{ p^i p_i}.
\end{equation}
In the one dimensional case this 
corresponds to the uncertainty relation given by $\Delta x \Delta p = 
[1 - 2 \alpha <p> + 4 \alpha^2 <p^2> ]$. These imply the existence of a minimum length  
$
 \Delta x \geq \Delta x_{min} \geq \alpha_0 \ell_{Pl}$, and a 
 a maximum 
momentum 
$
 \Delta p \leq \Delta p_{max} \leq \alpha_0^{-1} M_{Pl} c
$. 
Now the momentum in the coordinate representation can be written as    $p_i = \tilde p_i 
(1 - \alpha ||\tilde p||  + 2\alpha^2 ||\tilde p||^2)$, where $[x^i, \tilde p_j] = i \delta^i_j$, and so,  $ \tilde p_i = - \partial_i$. 
Thus, we can write 
\begin{equation}
p_i = -i\left(1 + \alpha \sqrt{-\partial^j \partial_j} - 2\alpha^2 \partial^j 
\partial_j\right)\partial_i.
\end{equation}

Now we will analyse deform the second quantized commutator similar to this deformation 
of the first quantized commutator. 
Thus,  we can write the commutator of a scalar field theory as 
\begin{equation}
 [\phi (x ), \pi (y)] = i \delta( x- y ) + i \alpha \mathcal{A} (x, y))  + i \alpha^2 \mathcal{B} (x-y), 
\end{equation}
where 
\begin{eqnarray}
 \mathcal{A} (x, y) &=& || \pi||\delta(x-y) + ||\pi||^{-1} \pi (x) \pi (y),  \nonumber \\
 \mathcal{B} (x, y) &=& ||\pi||^2 \delta(x-y) + 3 \pi(x) \pi (y). 
\end{eqnarray}
Here we have defined the norm of $||\pi||$ as follows, 
\begin{equation}
 || \pi || = \sqrt{\int dx \delta(x-y)  \pi(x) \pi (y) }. 
\end{equation}
This corresponds to taking the deformation for $\pi (x) $
\begin{equation}
 \pi (x) =  \left( 1 -  \alpha   || \tilde \pi || + 2 \alpha^2 || \tilde \pi||^2 \right)
\tilde \pi (x). 
\end{equation}
where 
\begin{equation}
  [\phi (x ), \tilde \pi (y)] = i \delta( x- y ).
\end{equation}
So, we have 
\begin{equation}
\tilde \pi (x) = -i \frac{\delta }{\delta \phi (x)}. 
\end{equation}
Thus,  we get 
\begin{eqnarray}
  \pi (x)&=&  -i \left( 1 +  \alpha   \sqrt{- \int dz dy\delta(z-y)  \frac{\delta }{\delta \phi (z)} \frac{\delta }{\delta \phi (y)} } 
 \right. \nonumber \\ && \left.  -  2 \alpha^2 \int dz dy\delta(z-y)  \frac{\delta }{\delta \phi (z)} \frac{\delta }{\delta \phi (y)} \right)
\frac{\delta }{\delta \phi (x)}.
\end{eqnarray}

Thus, we observe that the deformation of the second quantized canonical commutation relation 
induces non-locality in the quantum field theory. 
\section{Deformed  Wheeler-DeWitt Equation}
In this section, we will deform the Wheeler-DeWitt equation and analyse its consequences. 
The 
 line element in the Arnowitt-Deser-Misner $3+1$ decomposition of general relativity is given by 
\begin{equation}
ds^{2}=g_{\mu\nu}\left(  x\right)  dx^{\mu}dx^{\nu}=\left(  -N^{2}+N_{i}
N^{i}\right)  dt^{2}+2N_{j}dtdx^{j}+h_{ij}dx^{i}dx^{j}.
\end{equation}
where $N_{i}$ the shift function and $N$ is the lapse function. 
Thus, the Lagrangian for gravity in the Arnowitt-Deser-Misner $3+1$ decomposition of general relativity, can be written as 
\begin{equation}
\mathcal{L}\left[  N,N_{i},h_{ij}\right]  =\sqrt{-g}R=\frac{N{}\,\sqrt{^{3}h}}{2\kappa}
\left[  K_{ij}K^{ij}-K^{2}+\,\left(  ^{3}R-2\Lambda\right)  \right]  ,
\end{equation}
Here $\Lambda$ is the cosmological constant,  $^{3}R$ is the three dimensional scalar curvature, 
$K_{ij}$ is the second fundamental form, and 
 $K=$ $h^{ij}K_{ij}$ is the
trace of the second fundamental form. 
The Hamiltonian is obtained via Legendre transformation, 
\begin{equation}
\tilde H=
dx\left[  N {H} + N_{i}{H}^{i}\right]  ,
\end{equation}
where
\begin{eqnarray}
{H}&=& \left(  2\kappa\right)  G_{ijkl}\pi^{ij}\pi^{kl}-\frac{\sqrt
{^{3}h}}{2\kappa}\left(  ^{3}R-2\Lambda\right), 
\nonumber \\
{H}^{i}&=& -2\nabla_{j}\pi^{ji}, 
\end{eqnarray}
where $\pi^{ij}$ is the momentum conjugate to $h_{ij}$. 
Here $G_{ijkl}$ is 
defined by 
\begin{equation}
G_{ijkl}=\frac{1}{2\sqrt{h}}(h_{ik}h_{jl}+h_{il}h_{jk}-h_{ij}h_{kl}).
\end{equation}

The two classical constraints ${H}=0$,  and ${H}^{i}=0$, are obtained through the equation of motion. 
At the quantum level the constraints ${H}=0$, becomes the Wheeler-DeWitt equation, $\mathcal{H} \psi [h]=0$. Here  $\psi [h]$ is the wave function of the universe.
We will now use the modified canonical commutation relation to 
obtain a deformed version of this Wheeler-DeWitt equation. 
The  modified  canonical commutation relation are 
\begin{eqnarray}
 [h_{ij} (x), \pi^{kl} (y)] &=&  
 (\delta_i^k\delta_j^l + \delta_i^l \delta_j^k) (i \delta( x- y ) + i \alpha \mathcal{A} (x, y))  \nonumber \\ && 
 + i \alpha^2 \mathcal{B} (x-y)),  
\end{eqnarray}
where 
\begin{eqnarray}
 \mathcal{A} (x, y) &=& || \pi||\delta(x-y) + ||\pi||^{-1} \pi (x) \pi (y),  \nonumber \\
 \mathcal{B} (x, y) &=& ||\pi||^2 \delta(x-y) + 3 \pi(x) \pi (y). 
\end{eqnarray}
Here we have defined the norm of $||\pi||$ as follows, 
\begin{equation}
 || \pi || = \sqrt{\int dx dy \delta(x-y)  G_{ijkl} (x, y)\pi^{ij}(x) \pi^{kl} (y) }. 
\end{equation}
The momentum operator corresponding to this deformed canonical commutation relation is given by 
This corresponds to taking the deformation for $\pi (x) $
\begin{equation}
 \pi_{ij} (x) = \left( 1 -  \alpha   || \tilde \pi || + 2 \alpha^2 || \tilde \pi||^2 \right) 
\tilde \pi_{ij} (x). 
\end{equation}
where 
\begin{equation}
 [h_{ij} (x), \tilde \pi^{kl} (y)]  = i (\delta_i^k\delta_j^l + \delta_i^l \delta_j^k) \delta( x- y ). 
 \end{equation}
So, we have 
\begin{equation}
\tilde \pi_{ij} (x) = -i \frac{\delta }{\delta h_{ij} (x)}. 
\end{equation}
Thus,  we get 
\begin{eqnarray}
  \pi_{ij} (x)&=&  -i \left( 1 +  \alpha   \sqrt{- \int dzdy \delta(z-y) G_{klnm} (z, y) \frac{\delta }{\delta h_{kl} (z)} \frac{\delta }{\delta h_{mn} (y)} } 
 \right. \nonumber \\ && \left.  -  2 \alpha^2 \int dz dy\delta(z-y) G_{klmn} (z, y) \frac{\delta }{\delta h_{kl} (z)} \frac{\delta }{\delta h_{mn} (y)} \right)
 \nonumber \\ &&  \times \frac{\delta }{\delta h_{ij} (x)}.
\end{eqnarray}
Now we can write the deformed  Wheeler-DeWitt equation
\begin{equation}
  \mathcal{H}\psi\left[ h\right]  = 0
\end{equation}
where 
\begin{equation}
  \mathcal{H} = - \left(  2\kappa\right)  
  G_{ijkl}  \mathcal{D}
 \frac{\delta}{\delta h_{ij}}
  \mathcal{D}  
 \frac{\delta}{\delta h_{kl} }
-\frac{\sqrt
{^{3}h}}{2\kappa}\left(  ^{3}R-2\Lambda\right), 
\end{equation}
and 
\begin{eqnarray}
 \mathcal{D} &=& 1 +  \alpha \sqrt{-\int dx dy \delta(x-y)  G_{ijkl} (x, y) 
 \frac{\delta }{\delta h_{ij} (x)} \frac{\delta }{\delta h_{kl} (y)} } 
 \nonumber \\&& - 2\alpha^2 \int dx dy \delta(x-y)  G_{ijkl} (x, y)\frac{\delta }{\delta h_{ij} (x)} 
 \frac{\delta }{\delta h_{kl} (y)}. 
\end{eqnarray}

Now we consider a minisuperspace approximation to the  Wheeler-DeWitt equation. So, will will now consider a
closed universe filled with a vacuum of constant
energy density and the radiation, $\rho(a)=\rho_v+ \epsilon/a^4,$ where $\rho_v$ is the
vacuum energy density, $a$ is the scale factor and $\epsilon$ is a
constant characterizing the amount of radiation. 
The   
Friedman-Robertson-Walker metric for $k=1$ is given by 
\begin{equation}
ds^{2}=-N^{2}dt^{2}+a^{2}\left(  t\right)  d\Omega_{3}^{2},
\end{equation} 
where $d\Omega_{3}^{2}$ is the usual line element on the three sphere. 
Thus, we obtain the following result, 
\begin{equation}
\label{Fri}-\frac{3\pi}{4G}a\dot{a}^2-\frac{3\pi}{4G}a+2\pi^2a^3\rho(a)=0.
\end{equation}
If we assume $256\,\pi^2G^2\rho_v\,\epsilon/9<1$, then  a
big bang occurs at  $a=0$ and the universe expands to a maximum radius before tunneling 
into a phase of unbounded expansion.
Now we can write the following Lagrangian for this system 
\begin{equation}
  \mathcal{L}=-\frac{3\pi}{4G}a\dot{a}^2+\frac{3\pi a}{4G}-2\pi^2a^3\rho(a).
\end{equation}
We  can obtain a Hamiltonian from this Lagrangian as follows, 
\begin{equation}
H =-\frac{G}{3\pi}\frac{p^2}{a}-\frac{3\pi}{4G}a+2\pi^2a^3\rho(a).
\end{equation}
Now we deform the momentum operator for this minisuperspace model as
\begin{equation}
p = -i\left(1 + i \alpha \frac{d}{da} - 2\alpha^2 \frac{d^2}{d^2a}\right)\frac{d}{da}.
\end{equation}
The Wheeler-DeWitt equation corresponding to this representation of the momentum is given by 
\begin{equation}
  \mathcal{H} \psi(a) =0, 
\end{equation}
where 
\begin{eqnarray}
   \mathcal{H}  &=&  \left(\frac{G }{3\pi} \frac{d^2 \psi}{dx^2} - 2\alpha i \frac{G }{3\pi}
   \frac{d^3\psi}{dx^3} + 5 \alpha^2 \frac{G }{3\pi}
 \frac{d^4 \psi }{dx^4 } \right. \nonumber \\ && \left.  + \frac{3\pi}{4G}a^2+2\pi^2a^4\rho(a)\right).
\end{eqnarray}
This modified quantization is consistent with the following 
\begin{equation}
 \Delta a \Delta p = 1 - 2 \alpha <p> + 4 \alpha^2 <p^2> .
\end{equation}
 These imply the existence of a minimum length  
$
 \Delta a \geq \Delta a_{min} $.
It also implies the existence of a maximum 
momentum $
 \Delta p \leq \Delta p_{max}$. 
Thus, in the radius of the universe according to this modified Wheeler-DeWitt equation cannot shrink to zero. 
This way we can avoid the big bang singularity using this modified quantization. 
\section{Conclusion}
In this paper we generalized the deformation of first quantized commutating relations to   second quantized commutating relations.
Then we analysed the Wheeler-DeWitt equation using this formalism and demonstrated that in this formalism the big bang singularity is naturally avoided. 
It may be noted that  various other boundary conditions have been used for obtaining the 
 wave function of the universe. The wave function of the universe has also been constructed using a quantum tunneling transition
\cite{Vilenkin86}-\cite{Vilenkin82}. 
In fact, it is possible for a baby universe to be created by a quantum fluctuation of the vacuum. 
This universe can eventually jump into an inflationary period  and undergo a period of rapid expansion till 
its Hubble length becomes very large. This way a universe can be created by a quantum fluctuation of the vacuum. 
It will be interesting to study this mechanism for the creation of the universe using  
 the modified Wheeler-DeWitt equation. 

It may be noted that if a  single universe can be created from 
the quantum fluctuation of the vacuum, there is no reason why other universes cannot be similarly created. 
Thus, this model naturally predicts the existence of the multiverse. In fact, this model of inflation 
 is called the chaotic inflationary multiverse \cite{Linde86}. In this model, the 
total number of distinguishable locally Friedman universes generated by
eternal inflation is proportional to the exponent of the entropy of inflationary perturbations \cite{Linde}. 
The multiverse also appears naturally in the 
Multiverse can also be used to explain the landscape in the string
theory \cite{Davies04}. This is because all the $10^{500}$ different string theory vacuum states \cite{string}, can be viewed as 
 real vacuum states of different universes \cite{string0}. The  multiverse is most naturally analysed 
using a third quantized formalism of quantum gravity \cite{Strominger90}-\cite{third}. This is because the  Wheeler-DeWitt equation can be viewed as the 
 Schroedinger's equation for a single universe. Just as a single particle wave equation has to be second quantized to account for the creation and annihilation of 
particles, the Wheeler-DeWitt equation has to be third quantized to account for the 
creation and annihilation of universes.  It will be interesting to analyse the third quantization of this deformed Wheeler-DeWitt equation.


\begin{thebibliography}{99}
\bibitem{z4} M. Maggiore, Phys. Lett. B304, 65 (1993)
\bibitem{z5}M. I. Park, Phys. Lett. B659, 698 (2008) 
\bibitem{z2}D. Amati, M. Ciafaloni and G. Veneziano, Phys. Lett. B 216, 41 (1989)
\bibitem{zasaqsw}  A. Kempf, G. Mangano, and R. B. Mann, Phys. Rev. D 52, 1108 (1995) 
\bibitem{csdcas}L.  N.  Chang, D.  Minic, N. Okamura, and T.  Takeuchi, Phys.Rev. D65,  125027 (2002)
\bibitem{cscds}L.  N.  Chang, D.  Minic, N. Okamura, and T.  Takeuchi, Phys. Rev. D65, 125028 (2002) 
\bibitem{2z}S. Benczik, L. N.  Chang, D.  Minic, N.  Okamura, S.  Rayyan, and T.  Takeuchi,  Phys. Rev. D66, 026003 (2002)
\bibitem{z1}P.  Dzierzak, J.  Jezierski, P.  Malkiewicz, and W. Piechocki,  Acta Phys. Polon. B41, 717 (2010) 
\bibitem{1}     D. Amati, M. Ciafaloni, and G. Veneziano, Phys. Lett. B  216, 41  (1989) 
\bibitem{11}    M. Maggiore, Phys. Lett. B 304, 65 (1993)
\bibitem{12}    M. Maggiore,  Phys. Rev. D   49, 5182  (1994)
\bibitem{13}    M. Maggiore, Phys. Lett. B   319,  83 (1993)
\bibitem{14}    L. J. Garay,  Int. J. Mod. Phys. A   10,  145 (1995)
\bibitem{15}    F. Scardigli, Phys. Lett. B   452, 39  (1999)
\bibitem{17}    C. Bambi, F. R. Urban,  Class. Quantum Grav.   25,  095006 (2008)
\bibitem{18}    K. Nozari,  Phys. Lett. B.   629,  41 (2005) 
\bibitem{19}    K. Nozari, T. Azizi,  Gen. Relativ. Gravit.   38, 735  (2006)
\bibitem{10}    P. Pedram,  Int. J. Mod. Phys. D   19,  2003 (2010)
\bibitem{5}     A. Kempf, G. Mangano, and R. B. Mann,  Phys. Rev. D   52,  1108 (1995)
\bibitem{51}    A. Kempf, J. Phys. A   30, 2093 (1997) 
\bibitem{52}    F. Brau,  J. Phys. A   32, 7691 (1999) 
\bibitem{53}    K. Nozari, and B. Fazlpour, Chaos, Solitons and Fractals,   34, 224 (2007) 
\bibitem{54}    S. Das, and E. C. Vagenas,  Phys. Rev. Lett.   101,  221301 (2008)
\bibitem{n9}M.  Kober, Phys. Rev. D82, 085017 (2010)
\bibitem{skdj}V. Husain, D. Kothawala and S. S. Seahra, Phys. Rev. D 87, 025014 (2013) 
\bibitem{n}M. Kober, Int. J. Mod. Phys. A 26,   4251 (2011)
\bibitem{2}     J. Magueijo, and L. Smolin,  Phys. Rev. Lett.   88,  190403 (2002)
\bibitem{21}    J. Magueijo, and L. Smolin, Phys. Rev. D   71,  026010 (2005)
\bibitem{3}     J. L. Cortes, and J. Gamboa,  Phys. Rev. D   71,  065015 (2005)
\bibitem{n1} J. Magueijo,  and L. Smolin, Class. Quant. Grav. 21,  1725 (2004)
\bibitem{n2} J. J.  Peng, and S. Q. Wu,  Gen. Rel. Grav. 40, 2619 (2008) 
\bibitem{main2} S. Das, E. C. Vagenas, and A. F. Ali,  Phys. Lett. B   690,  407 (2010)
 \bibitem{n4}A. F. Ali, S. Das and E. C. Vagenas, Phys. Lett. B 678, 497 (2009) 
 \bibitem{fwrw} A. F. Ali and B. Majumder, arXiv:1402.5104
 \bibitem{DeWitt67} B. S. DeWitt, Phys. Rev. 160, 1113 (1967)
\bibitem{Wheeler57} J. A. Wheeler,  Ann. Phys. 2, 604 (1957)
\bibitem{Hartle83} J. B. Hartle and S. W. Hawking, Phys. Rev. D 28,   2960 (1983)
\bibitem{Hawking} S. W. Hawking, T.  Hertog and H.  Reall,   Phys.  Rev.  D 62,  043501 (2000)
\bibitem{Hawking81} S. W. Hawking, Pontif. Acad. Sci. Scrivaria 48, 563 (1982)
 \bibitem{n5}    P. Pedram,  Europhys. Lett.   89, 50008 (2010) 
\bibitem{n6} P. Pedram, K. Nozari, and S. H. Taheri, JHEP. 1103, 093 (2011) 
\bibitem{n7} A. F.  Ali, S.  Das, and E.  C. Vagenas, Phys. Rev. D84, 044013 (2011)
\bibitem{Vilenkin86} A. Vilenkin, Phys. Rev. D 33, 3560 (1986)
\bibitem{Vilenkin82} A. Vilenkin, Phys. Lett. B 117,  25 (1982)
\bibitem{Linde86} A. Linde, Mod. Phys. Lett A 1, 81 (1986)
\bibitem{Linde}A. Linde and V. Vanchurin, Phys. Rev. D81, 083525 (2010)
\bibitem{Davies04} S. K. Ashok and M. R. Douglas, JHEP 01, 060 (2004)
\bibitem{string}F. Denef and M. R. Douglas, JHEP 0405, 072 (2004)
\bibitem{string0} L. M. Houghton and R. Holman, JCAP 0902, 006 (2009)
\bibitem{Strominger90} A. Strominger, Nucl.  Phy. B 321,  481 (1989)
\bibitem{m}M. Faizal, Mod. Phys. Lett. A27,  1250007 (2012) 
\bibitem{third}Y.  Ohkuwa and Y.  Ezawa, Class. Quant. Grav. 29, 215004  (2012) 
\bibitem{ia}Y.  Ohkuwa, Int. J. Mod. Phys. A 13, 4091 (1998)
\end{thebibliography}
\end{document}